\documentclass[10pt,prl,twocolumn,showpacs]{revtex4}
\usepackage{graphicx,amsmath,amssymb}
\begin{document}

\renewcommand{\Re}{\mathop{\mathrm{Re}}}
\renewcommand{\Im}{\mathop{\mathrm{Im}}}
\renewcommand{\b}[1]{\mathbf{#1}}
\renewcommand{\c}[1]{\mathcal{#1}}
\renewcommand{\u}{\uparrow}
\renewcommand{\d}{\downarrow}
\newcommand{\bsigma}{\boldsymbol{\sigma}}
\newcommand{\blambda}{\boldsymbol{\lambda}}
\newcommand{\tr}{\mathop{\mathrm{tr}}}
\newcommand{\sgn}{\mathop{\mathrm{sgn}}}
\newcommand{\sech}{\mathop{\mathrm{sech}}}
\newcommand{\diag}{\mathop{\mathrm{diag}}}
\newcommand{\half}{{\textstyle\frac{1}{2}}}
\newcommand{\sh}{{\textstyle{\frac{1}{2}}}}
\newcommand{\ish}{{\textstyle{\frac{i}{2}}}}
\newcommand{\thf}{{\textstyle{\frac{3}{2}}}}

\title{Topological quantization in units of the fine structure constant}

\author{Joseph Maciejko$^{1,2}$, Xiao-Liang Qi$^{3,1,2}$, H. Dennis Drew$^4$, and Shou-Cheng Zhang$^{1,2}$}

\affiliation{$^1$Department of Physics, Stanford
University, Stanford, CA 94305, USA\\
$^2$Stanford Institute for Materials and Energy Sciences, SLAC
National Accelerator Laboratory, Menlo Park, CA 94025, USA\\
$^3$Microsoft Research, Station Q, Elings Hall, University of
California, Santa Barbara, CA 93106, USA\\
$^4$Center for Nanophysics and Advanced Materials, Department of
Physics, University of Maryland, College Park, MD 20742, USA}
\date\today

\begin{abstract}
Fundamental topological phenomena in condensed matter physics are
associated with a quantized electromagnetic response in units of
fundamental constants. Recently, it has been predicted
theoretically that the time-reversal invariant topological
insulator in three dimensions exhibits a topological
magnetoelectric effect quantized in units of the fine structure
constant $\alpha=e^2/\hbar c$. In this Letter, we propose an optical
experiment to directly measure this topological quantization
phenomenon, independent of material details. Our proposal also
provides a way to measure the half-quantized Hall conductances on
the two surfaces of the topological insulator independently of
each other.
\end{abstract}


\maketitle

Topological phenomena in condensed matter physics are typically
characterized by the exact quantization of the electromagnetic
response in units of fundamental constants. In a superconductor
(SC), the magnetic flux is quantized in units of the flux quantum
$\phi_0\equiv\frac{h}{2e}$; in the quantum Hall effect (QHE), the
Hall conductance is quantized in units of the conductance quantum
$G_0\equiv\frac{e^2}{h}$. Not only are these fundamental physical
phenomena, they also provide the most precise metrological
definition of basic physical constants. For instance, the
Josephson effect in SC allows the most precise measurement of the
flux quantum which, combined with the measurement of the quantized
Hall conductance, provides the most accurate determination of
Planck's constant $h$ to date~\cite{Mohr2008}. The remarkable
observation of such precise quantization phenomena in these
imprecise, macroscopic condensed matter systems can be understood
from the fact that they are described in the low-energy limit by
topological field theories (TFT) with quantized coefficients. For
instance, the QHE is described by the topological Chern-Simons
theory~\cite{Zhang1992} in $2+1$ dimensions, with coefficient
given by the quantized Hall conductance. SC can be described by
the topological $BF$ theory~\cite{Hansson2004} with coefficient
corresponding to the flux quantum.

More recently, a new topological state in condensed matter
physics, the time-reversal (${\cal T}$) invariant topological
insulator (TI), has been investigated
extensively~\cite{Qi2010,Moore2009,Hasan2010}. The concept of TI
can be defined most generally in terms of the TFT~\cite{Qi2008}
with effective Lagrangian
\begin{align}\label{Laxion}
\mathcal{L}=\frac{1}{8\pi}\left(\varepsilon\b{E}^2-\frac{1}{\mu}\b{B}^2\right)
+\frac{\theta}{2\pi}\frac{\alpha}{2\pi}\b{E}\cdot\b{B},
\end{align}
where $\b{E}$ and $\b{B}$ are the electromagnetic fields,
$\varepsilon$ and $\mu$ are the dielectric constant and magnetic
permeability, respectively, and $\theta$ is an angular variable
known in particle physics as the axion angle~\cite{Wilczek1987}.
Under periodic boundary conditions, the partition function and all
physical quantities are invariant under shifts of $\theta$ by any
multiple of $2\pi$. Since $\b{E}\cdot\b{B}$ is odd under ${\cal
T}$, the only values of $\theta$ allowed by ${\cal T}$ are $0$ or
$\pi$ (modulo $2\pi$). The second term of Eq.~(\ref{Laxion}) thus
defines a TFT with coefficient quantized in units of the fine
structure constant $\alpha\equiv\frac{e^2}{\hbar c}$. The TFT is
generally valid for interacting systems, and describes a quantized
magnetoelectric response denoted topological magnetoelectric
effect (TME)~\cite{Qi2008}. The quantization of the axion angle
$\theta$ depends only on the ${\cal T}$ symmetry and the bulk
topology; it is therefore universal and independent of any
material details. More recently, it has been shown~\cite{Wang2009}
that the TFT~\cite{Qi2008} reduces to the topological band theory
(TBT)~\cite{Kane2005,Fu2007,Moore2007} in the noninteracting
limit. Interestingly, the TME is the first topological
quantization phenomenon in units of $\alpha$. It can therefore be
combined with the two other known topological phenomena in
condensed matter, the QHE and SC, to provide a metrological
definition of the three basic physical constants, $e$, $h$, and
$c$.

The TME has not yet been observed experimentally. An insight into
why this is so can be gained by comparing the $3+1$ dimensional
TFT~(\ref{Laxion}) of TI to the $2+1$ dimensional Chern-Simons TFT
of the QHE~\cite{Zhang1992}. In $2+1$ dimensions, the topological
Chern-Simons term is the only term which dominates the long-wavelength behavior of the system, which leads to the universal quantization of the Hall conductance. On the other hand,
in $3+1$ dimensions the topological $\theta$-term in
Eq.~(\ref{Laxion}) and the Maxwell
term are equally important in the long wavelength limit. Therefore, one has to be careful when designing an
experiment to observe the topological quantization of the TME, in
which the dependence on the non-topological materials constants
$\varepsilon$ and $\mu$ are removed.

\begin{figure}[t]
\begin{center}
\includegraphics[width=0.9\columnwidth]{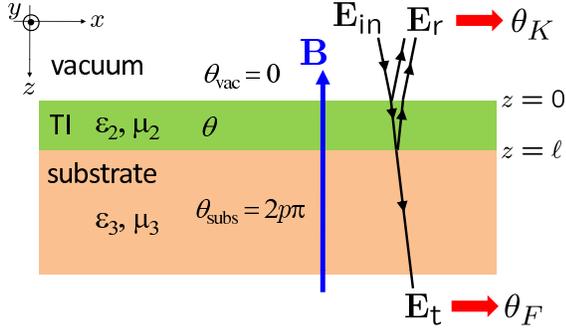}
\end{center}
\caption{(color online). Measurement of Kerr and Faraday angles for
a TI thick film of thickness $\ell$ and optical constants
$\varepsilon_2,\mu_2$ on a topologically trivial insulating
substrate with optical constants $\varepsilon_3,\mu_3$, in a
perpendicular magnetic field $\b{B}$. (We consider normal incidence
in the actual proposal but draw light rays with a finite incidence
angle in the figure for clarity.) The external magnetic field can be
replaced by a thin magnetic coating on both TI surfaces, as
suggested in Ref.~\cite{Qi2008}.} \label{fig:fig1}
\end{figure}

In this Letter, we propose an optical experiment to observe the
topological quantization of the TME in units of $\alpha$,
\emph{independent of material properties of the TI} such as
$\varepsilon$ and $\mu$. This experiment could be performed
on any of the available TI materials, such as the Bi$_2$Se$_3$, Bi$_2$Te$_3$,
Sb$_2$Te$_3$ family or the recently discovered thallium-based compounds~\cite{chalcogenides}. Consider a TI thick film of thickness
$\ell$ with optical constants $\varepsilon_2,\mu_2$ and axion
angle $\theta$ deposited on a topologically trivial insulating
substrate with optical constants
$\varepsilon_3,\mu_3$~(Fig.~\ref{fig:fig1}).
The vacuum outside the TI has $\varepsilon=\mu=1$ and trivial
axion angle $\theta_\mathrm{vac}=0$. The substrate being also
topologically trivial, both interfaces at $z=0$ and $z=\ell$
support a domain wall of $\theta$ giving rise to a surface QHE
with half-quantized surface Hall conductance
$\sigma_H^s=(n+\frac{1}{2})\frac{e^2}{h}$ with
$n\in\mathbb{Z}$~\cite{Qi2008}. The factor of $\frac{1}{2}$ is a
topological property of the bulk and is protected by the ${\cal
T}$ symmetry. On the other hand, the value of $n$ depends on the
details of the interface and may thus be different for the two
interfaces. To account for this general case we assign
$\theta_\mathrm{subs}=2p\pi$ with $p\in\mathbb{Z}$ to the
topologically trivial substrate, corresponding to
$\sigma_H^{s,0}=\frac{\theta}{2\pi}\frac{e^2}{h}$ on the $z=0$
interface and
$\sigma_H^{s,\ell}=(p-\frac{\theta}{2\pi})\frac{e^2}{h}$ on the
$z=\ell$ interface. The experiment consists in shining normally
incident monochromatic light with frequency $\omega$ on the TI
film, and measuring the Kerr angle $\theta_K$ of the reflected
light and Faraday angle $\theta_F$ of the transmitted light.
However, the effective theory~(\ref{Laxion}) applies only in the regime $\omega\ll E_g/\hbar$ where $E_g$ is the
surface gap~\cite{Qi2008}. Such a surface gap can be opened by a
thin magnetic coating on both surfaces of the TI, as first
suggested in Ref.~\cite{Qi2008}, or by an applied perpendicular
magnetic field $\b{B}=B\hat{\b{z}}$ (Fig.~\ref{fig:fig1}) through
the surface Zeeman effect as well as the exchange coupling between
surface electrons and the paramagnetic bulk. We
discuss the experimentally simpler case of the external magnetic field. For
incident light linearly polarized in the $x$ direction
$\b{E}_\mathrm{in}=E_\mathrm{in}\hat{\b{x}}$, the Kerr and Faraday
angles are defined by $\tan\theta_K=E_\mathrm{r}^y/E_\mathrm{r}^x$
and $\tan\theta_F=E_\mathrm{t}^y/E_\mathrm{t}^x$, respectively,
with
$\b{E}_\mathrm{r}=E_\mathrm{r}^x(-\hat{\b{x}})+E_\mathrm{r}^y\hat{\b{y}}$
and
$\b{E}_\mathrm{t}=E_\mathrm{t}^x\hat{\b{x}}+E_\mathrm{t}^y\hat{\b{y}}$
the reflected and transmitted electric fields, respectively
(Fig.~\ref{fig:fig1}). Furthermore, $\theta_K$ and $\theta_F$ are
to be measured as a function of $B$. The angles that we discuss in
the following are defined as the linear extrapolation
of $\theta_K(B)$ and $\theta_F(B)$ as $B\rightarrow 0^+$, in which
limit the non-topological bulk contribution to optical rotation is
removed~\cite{Qi2008}.

The problem of optical rotation at a TI/trivial insulator
interface has been studied
before~\cite{Qi2008,Karch2009,Chang2009}. In general, $\theta_K$
and $\theta_F$ depend on the optical constants
$\varepsilon_2,\mu_2$ of the TI. In the thick film geometry
considered here, they will also depend in a complicated manner on
the optical constants $\varepsilon_3,\mu_3$ of the substrate, the
film thickness $\ell$, and the photon frequency $\omega$, due to
multiple reflection effects at the two interfaces. It seems
therefore dubious that one could extract the exact quantization of
the TME from such a measurement. However, we find that these multiple reflection effects can
be used for a universal measurement of the TME, with no explicit
dependence on $\varepsilon_2,\mu_2,\varepsilon_3,\mu_3,\ell$, and
$\omega$.
\begin{figure}[t]
\begin{center}
\includegraphics[width=0.9\columnwidth]{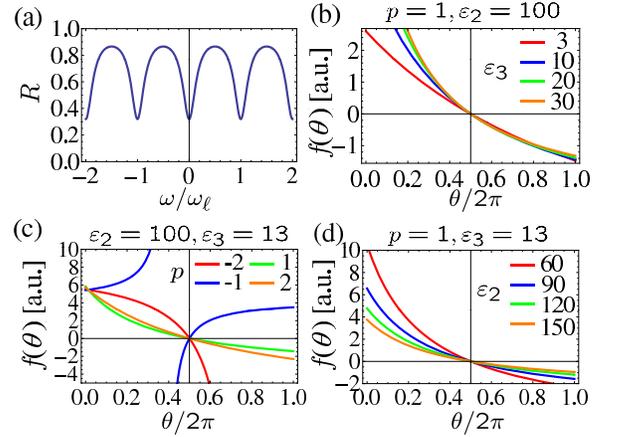}
\end{center}
\caption{(color online). (a) Reflectivity $R$ as a function of photon frequency $\omega$
in units of the characteristic frequency $\omega_\ell$ for a topological
insulator Bi$_2$Se$_3$ thick film on a Si substrate; universal function
$f(\theta)$ for different values of (b) the substrate dielectric
constant $\varepsilon_3$, (c) $p$, the total surface Hall conductance
in units of $\frac{e^2}{h}$, and (d) the TI dielectric constant
$\varepsilon_2$. The position of the zero crossing is universal and
provides an experimental demonstration of the quantized TME.}
\label{fig:fig2}
\end{figure}

In Fig.~\ref{fig:fig2}(a) we plot the reflectivity
$R\equiv|\b{E}_\mathrm{r}|^2/|\b{E}_\mathrm{in}|^2$ as a function
of photon frequency $\omega$ in units of a characteristic
frequency
$\omega_\ell\equiv\frac{c}{\sqrt{\varepsilon_2\mu_2}}\frac{\pi}{\ell}$,
for $\varepsilon_2=100$, $\varepsilon_3=13$, and $\mu_2=\mu_3=1$,
appropriate for a topological Bi$_2$Se$_3$~\cite{Zhang2009} thin
film on a Si substrate~\cite{laforge2009,butch2010,JasonHancock}.
We observe that minima in $R$ occur when $\omega/\omega_\ell$ is an integer, corresponding to $\ell$ being an
integer multiple of $\frac{\lambda_2}{2}$ with
$\lambda_2=\frac{2\pi c}{\omega\sqrt{\varepsilon_2\mu_2}}$ the
photon wavelength inside the TI. For radiation in the terahertz range this corresponds to $\ell\sim 100$~$\mu$m. When $\omega$ is tuned to any of
these minima, we find
\begin{align}\label{minima1}
\tan\theta_K'=\frac{4\alpha p}{Y_3^2-1+4\alpha^2p^2},
\hspace{5mm}
\tan\theta_F'=\frac{2\alpha p}{Y_3+1},
\end{align}
where $Y_i\equiv\sqrt{\varepsilon_i/\mu_i}$ is the admittance of
region $i$, and the prime indicates rotation angles
measured at a reflectivity minimum, i.e. for
$\omega/\omega_\ell\in\mathbb{Z}$. We see that $\theta_K'$ and
$\theta_F'$ are independent of the TI optical constants
$\varepsilon_2,\mu_2$. Equation~(\ref{minima1}) corresponds simply
to the results of Ref.~[\onlinecite{Qi2008,Karch2009}] for a
\emph{unique} interface with axion domain wall
$\Delta\theta=2p\pi$. Moreover, the two angles can be
combined~\cite{JiangZhe} to obtain a universal result independent
of both TI $\varepsilon_2,\mu_2$ and substrate
$\varepsilon_3,\mu_3$ properties,
\begin{align}\label{pa}
\frac{\cot\theta_F'+\cot\theta_K'}{1+\cot^2\theta_F'}=\alpha p,
\hspace{5mm}p\in\mathbb{Z}.
\end{align}
Since the rotation angles are measured at a reflectivity
minimum, Eq.~(\ref{pa}) has no explicit dependence on $\ell$ or
$\omega$ either. Equation~(\ref{pa}) clearly expresses the
topological quantization in units of $\alpha$ solely in terms of
experimentally measurable quantities, and is the first important
result of this work.

However, neither Eq.~(\ref{minima1}) nor Eq.~(\ref{pa}) depend
explicitly on the TI axion angle $\theta$, and one may ask whether
Eq.~(\ref{pa}) is at all an indication of nontrivial bulk
topology. In fact, Eq.~(\ref{pa}) describes the topological
quantization of the \emph{total} Hall conductance of both surfaces
$\sigma_H^{s,\mathrm{tot}}=\sigma_H^{s,0}+\sigma_H^{s,\ell}=p\frac{e^2}{h}$,
which holds independently of possible $\mathcal{T}$ breaking in the bulk. In the
special case that the two surfaces have the same surface Hall conductance, we have
$p=2\sigma_H^{s,0}=\frac{\theta}{\pi}$ and Eq.~(\ref{pa}) is
sufficient to determine the bulk axion angle $\theta$. However,
for a TI film on a substrate the two surfaces are generically
different and can have different Hall conductance. To obtain the
axion angle $\theta$ in the more general case
of different surfaces, we propose another optical measurement performed at reflectivity
\emph{maxima} $\omega=(n+\frac{1}{2})\omega_l$, $n\in\mathbb{Z}$
[Fig.~\ref{fig:fig2}(a)]. We denote by $\theta_K''$ and
$\theta_F''$ the Kerr and Faraday angles measured at an arbitrary
reflectivity maximum. In contrast to $\theta_K'$ and $\theta_F'$
[Eq.~(\ref{minima1})], these depend on $\varepsilon_2,\mu_2$ as
well as on $\varepsilon_3,\mu_3$,
\begin{align}\label{maxima}
\tan\theta_K''&=\frac{4\alpha
\left[Y_2^2\left(p-\frac{\theta}{2\pi}\right)
-\tilde{Y}_3^2\frac{\theta}{2\pi}\right]}
{\tilde{Y}_3^2-Y_2^4+4\alpha^2
\left[2Y_2^2\frac{\theta}{2\pi}\left(p-\frac{\theta}{2\pi}\right)
-\tilde{Y}_3^2\left(\frac{\theta}{2\pi}\right)^2\right]},
\nonumber\\
\tan\theta_F''&=\frac{2\alpha
\left(p-\frac{\theta}{2\pi}+Y_3\frac{\theta}{2\pi}\right)}
{Y_3+Y_2^2-4\alpha^2\frac{\theta}{2\pi}
\left(p-\frac{\theta}{2\pi}\right)},
\end{align}
where we define
$\tilde{Y}_3^2=Y_3^2+4\alpha^2\left(p-\frac{\theta}{2\pi}\right)^2$.
More importantly, $\theta_K''$ and $\theta_F''$ depend explicitly
on the TI axion angle $\theta$. It is readily checked that
Eq.~(\ref{maxima}) reduces to Eq.~(\ref{minima1}) in the
single-interface limit $\theta=2p\pi$, $Y_2=Y_3$ or $\theta=0$,
$Y_2=1$. In general however, from the knowledge of $p$
[Eq.~(\ref{pa})] and either $\theta_K'$ or $\theta_F'$ we can
extract $Y_3$ by using
Eq.~(\ref{minima1}) without performing any separate measurement. Moreover, $\theta_K''$ and
$\theta_F''$ can be combined to cancel
the explicit dependence on the TI properties
$\varepsilon_2,\mu_2$. We solve for $Y_2^2$ in
Eq.~(\ref{maxima}) in terms of $\theta_F''$, say, and substitute
the resulting expression $Y_2^2=Y_2^2(\theta)$ into the equation
for $\theta_K''$ in Eq.~(\ref{maxima}). The result can be expressed in the form
$f(\theta_K',\theta_F',\theta_K'',\theta_F'';p,\theta)=0$ where
$f$ is `universal' in the sense that it does not depend explicitly
on any material parameter $\varepsilon_i,\mu_i$. Substituting the
experimental values of $\theta_K',\theta_F',\theta_K'',\theta_F''$
and $p$ into this expression, we obtain a function
of a single variable $f(\theta)$. If we plot
$f$ as a function of $\theta$, the zero crossing $f(\theta)=0$ gives the value of the bulk axion angle
$\theta$ with no $2\pi$ ambiguity. Plots of the universal function $f$ are given in
Fig.~\ref{fig:fig2}(b), (c), and (d) for different values of the
material parameters $\varepsilon_2,\varepsilon_3,p$ (setting
$\mu_2=\mu_3=1$ without loss of generality) and for a bulk axion
angle $\theta=\pi$. The zero crossing point is independent of
material parameters and, together with Eq.~(\ref{pa}), can provide
a universal experimental demonstration of the quantization of the
TME in the TI bulk. In a \emph{thin} film geometry
$\ell\ll\frac{\lambda_2}{2}$ corresponding to
$\omega\ll\omega_\ell$, the optical response is always given by
the sum of the Hall conductivities of the two surfaces. Therefore,
thick films $\ell\geq\frac{\lambda_2}{4}$ to allow destructive
interference and reflectivity maxima are essential to the
measurement of the bulk TME.

\begin{figure}[t]
\begin{center}
\includegraphics[width=0.9\columnwidth]{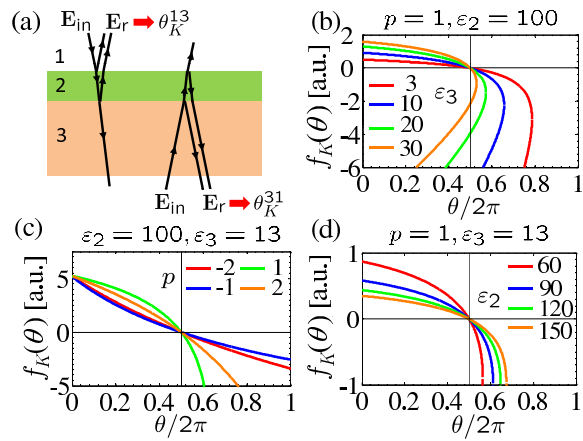}
\end{center}
\caption{(color online). (a) Kerr-only measurement setup, with material
parameters the same as indicated in Fig.~\ref{fig:fig1}; (b), (c)
and (d): universal function $f_K(\theta)$ for different material
parameters [same as in Fig.~\ref{fig:fig2}(b), (c), (d)]. As in
Fig.~\ref{fig:fig2}, the position of the zero crossing is universal and
provides an experimental demonstration of the quantized TME.}
\label{fig:fig3}
\end{figure}
Our proposal so far necessitates the measurement of both Kerr and
Faraday angles. We now show that it is possible to extract $p$ and
$\theta$ from Kerr measurements alone, if the Kerr angle is
measured in both directions [Fig.~\ref{fig:fig3}(a)]. Indeed,
while the Faraday angle is generally independent of the direction
of propagation~\cite{LL_EM}, the Kerr angle depends on it. Here we
exploit this asymmetry of the Kerr angle to extract $p$ and
$\theta$. We denote by $\theta_K^{\prime 13}$ and
$\theta_K^{\prime\prime 13}$ the Kerr angles defined previously in
Eq.~(\ref{minima1}) and (\ref{maxima}), respectively. Conversely,
we denote by $\theta_K^{\prime 31}$ and $\theta_K^{\prime\prime
31}$ the Kerr angles for light traveling in the opposite
direction, i.e. incident from the substrate
[Fig.~\ref{fig:fig3}(a)]. As before, the prime and double prime
correspond to angles measured at reflectivity minima and maxima,
respectively. We find
\begin{align}\label{thetaK31}
\tan\theta_K^{\prime 31}&=-\frac{4\alpha pY_3}{Y_3^2-1+4\alpha^2p^2},
\\
\tan\theta_K^{\prime\prime 31}&=
\frac{4\alpha Y_3\left[Y_2^2\frac{\theta}{2\pi}-
\gamma\left(p-\frac{\theta}{2\pi}\right)\right]}
{\gamma Y_3^2+4\gamma\alpha^2\left[p^2-\left(\frac{\theta}{2\pi}\right)^2
\right]-Y_2^4-8\alpha^2Y_2^2\left(\frac{\theta}{2\pi}\right)^2},
\nonumber
\end{align}
where we define
$\gamma\equiv1+4\alpha^2\left(\frac{\theta}{2\pi}\right)^2$. As
previously, $\theta_K^{\prime 13}$ and $\theta_K^{\prime 31}$ can
be combined to eliminate $Y_3$ and provide a universal measure of
$p\in\mathbb{Z}$,
\begin{align}\label{pKerr}
\cot\theta_K^{\prime 13}-\sgn p
\sqrt{1+\cot^2\theta_K^{\prime 13}(1-\tan^2\theta_K^{\prime 31})}
=2\alpha p,
\end{align}
provided $Y_3^2\equiv\varepsilon_3/\mu_3>1+4\alpha^2 p^2$, which
is satisfied in practice for low $p$ since $\alpha^2\sim
10^{-4}$. Furthermore, comparing Eq.~(\ref{thetaK31}) for
$\theta_K^{\prime 31}$ to Eq.~(\ref{minima1}) for
$\theta_K^{\prime 13}$ we see that $Y_3$ is easily obtained as $Y_3=-\cot\theta_K^{\prime 13}\tan\theta_K^{\prime 31}$. Finally, to extract the bulk axion angle $\theta$, we need to
solve for $Y_2^2$ in Eq.~(\ref{maxima}) in terms of
$\theta_K^{\prime\prime 13}$, and substitute the resulting
expression $Y_2^2=Y_2^2(\theta)$ into the equation for
$\theta_K^{\prime\prime 31}$ in Eq.~(\ref{thetaK31}). The result
of this analysis can once again be expressed in the form
$f_K(\theta_K^{\prime 13},\theta_K^{\prime
31},\theta_K^{\prime\prime 13},\theta_K^{\prime\prime
31};p,\theta)=0$, where $f_K$ is a `universal' function which only
depends on the measured Kerr angles. As before, we substitute into
$f_K$ the experimental values of $\theta_K^{\prime
13},\theta_K^{\prime 31},\theta_K^{\prime\prime
13},\theta_K^{\prime\prime 31}$ and $p$ [obtained from
Eq.~(\ref{pKerr})] and obtain a function of a single variable
$f_K(\theta)$ which crosses zero at the value of the bulk axion
angle with no $2\pi$ ambiguity. In Fig.~\ref{fig:fig3}(b), (c) and (d) we plot the universal
function $f_K$ for different values of the material parameters
$\varepsilon_2,\varepsilon_3,p$ and for a bulk axion angle
$\theta=\pi$. The zero crossing point is independent of material parameters and,
together with Eq.~(\ref{pKerr}), provides another means to
demonstrate experimentally the universal quantization of the TME
in the bulk of a TI.

Recent work~\cite{tse2010} has addressed the similar problem of
optical rotation on a TI film, and found interesting and
novel results for the rotation angles. However, these
results hold only in certain limits which are less general than
the ones discussed in this work. First, Ref.~\cite{tse2010}
considers a free-standing TI film in vacuum. Most films are grown
on a substrate which can affect the physics qualitatively. For
instance, the giant Kerr rotation
$\theta_K=\tan^{-1}(1/\alpha)\simeq\pi/2$ found in
Ref.~\cite{tse2010} is a special case of our Eq.~(\ref{minima1})
with $p=1$ and $\varepsilon_3/\mu_3=1$. It is dramatically
suppressed when $\varepsilon_3/\mu_3-1$ is greater than
$\alpha^2\sim 10^{-4}$, which is typically the case in practice.
Second, in Ref.~\cite{tse2010} a correction proportional to
$\Delta/\epsilon_c$ was introduced to the surface Hall
conductance, where $\Delta$ is the ${\cal T}$-breaking Dirac mass
and $\epsilon_c$ is a non-universal high-energy cutoff. According
to the general bulk TFT of the TI~\cite{Qi2008}, the surface Hall
conductance is always quantized as long as the surface is gapped
and the bulk is ${\cal T}$-invariant (in the $B\rightarrow 0$
limit). Thus we conclude that such a non-universal correction is
absent and the requirement $\Delta\ll\epsilon_c$ is not necessary
within the TFT approach~\cite{Qi2008}. This difference clearly
demonstrates the power of the TFT approach~\cite{Qi2008} in
predicting universally quantized topological effects in condensed
matter physics.

We acknowledge helpful discussions with A. Fried, T. L. Hughes, A.
Kapitulnik, R. Li, R. Maciejko, and especially with J. N. Hancock.
J.M. is supported by the Stanford Graduate Foundation. This work is supported
by the Department of Energy, Office of Basic Energy Sciences,
Division of Materials Sciences and Engineering, under contract
DE-AC02-76SF00515, and the NSF MRSEC at University of Maryland
under Grant No. DMR-0520471.

\bibliography{alpha}

\end{document}